
\input harvmac
\input epsf

\def\rhob{{\rho\kern-0.465em \rho}}

\def\ontopss#1#2#3#4{\raise#4ex \hbox{#1}\mkern-#3mu {#2}}

\setbox\strutbox=\hbox{\vrule height12pt depth5pt width0pt}

\def\strut{\relax\ifmmode\copy\strutbox\else\unhcopy\strutbox\fi}

\nref\rnm{Th. Niemeijer, Physica 36, (1967) 377.}  
\nref\rb{N. Bloembergen, Physica 15 (1949) 386.}
\nref\rfor{D. Forster, {\it Hydrodynamic Fluctuations, Broken Symmetry
and Correlation Functions}, (Benjamin, Reading MA, 1975)}
\nref\rsvw{M. Steiner, J. Villain and C.G. Windsor, Adv. Phys. 25
(1976) 87.}
\nref\rsl{A. Sur and I.J.Lowe, Phy. Rev. B11, (1975) 1980.}
\nref\rrmp{J.M.R. Roldan, B.M. McCoy and J.H.H. Perk, Physica 136A
(1986) 255.}
\nref\rbs{U.Brandt and J. Stolze, Z. Phys. B64 (1986) 327.}
\nref\bl{M. B{\"o}hm and H. Leshke, J. Phys. A 25 (1992) 1043.}
\nref\rfls{K. Fabricius, U. L{\" o}w and J. Stolze, Phys. Rev. B 55
(1997) 5833.}
\nref\rmc{B.M. McCoy, in {\it Statistical Mechanics and Field Theory},
ed. V.V. Bazhanov and C.J. Burden, (World Scientific 1995) 26.}
\nref\rczp{H. Castella, X. Zotos and P. Prelovsek,
Phys. Rev. Letts. 74 (1995) 972; X. Zotos and P. Prelovsek, Phys. Rev. B
53 (1996) 983.}
\nref\rnar{B.N. Narozhny, Phys. Rev. B 54 (1996) 3311; and to be published.}
\nref\rgma{G. M{\"u}ller, Phys. Rev. Letts. 60 (1988) 2785.}
\nref\rgl{R.W. Gerling and D.P Landau, Phys. Rev. Letts. 63 (1989)
812. }
\nref\rgmb{G. M{\"u}ller, Phys. Rev. Letts. 63 (1989) 813.}
\nref\rglb{R.W. Gerling and D.P Landau, Phys. Rev. B42 (1990) 8214.}
\Title{\vbox{\baselineskip12pt
  \hbox{ITPSB 97-32}}}
  {\vbox{\centerline{No spin diffusion in the spin 1/2 XXZ chain at $T=\infty:$}
        \centerline{Numerical asymptotics}}}


  \centerline{ Barry~M.~McCoy\foot{mccoy@max.physics.sunysb.edu}}               
  \bigskip\centerline{\it Institute for Theoretical Physics}
  \centerline{\it State University of New York}
  \centerline{\it Stony Brook,  NY 11794-3840}
  \bigskip
  \Date{\hfill 06/97}

  \eject

\centerline{\bf Abstract}
We  analyze  the recent numerical computations made by
Fabricius, L{\" o}w and Stolze to show that the long time behavior of
the zz correlation function of the spin 1/2 XXZ chain at $T=\infty$ 
is very well fit by the formula $t^{-d}[A+Be^{-\gamma (t-t_0)}\cos
\Omega(t-t_0)]$ where $d$ is substantially greater than 1/2. This
confirms the conclusion that there is no spin diffusion in this model.

\newsec{Introduction}

The spin 1/2 XXZ chain of $N$ site with periodic boundary conditions
specified by the Hamiltonian
\eqn\ham{H={1\over 2}\sum_{i=1}^{N}\left(\sigma^x_i \sigma^x_{i+1}+
\sigma^y_i \sigma^y_{i+1}+{\Delta}\sigma^x_i \sigma^x_{i+1}\right)}
where $\sigma^j_i$ is the $y=x,y,z$ Pauli spin matrix at site $i$
is well known to be an integrable system. As such it is to be expected
that all the spin correlation functions should be analytically
computable. In the past decade there has been much work done to
fulfill this promise but much still remains to be done. In particular
there has been no published work done on the time dependent
autocorrelation function 
 at infinite  temperature 
\eqn\defn{S(t;\Delta)=<\sigma^z_0(t)\sigma^z_0(0)>={\rm
lim}_{N\rightarrow \infty}2^{-N}Tr e^{-itH}\sigma^z_0 e^{itH}\sigma^z_0} 
other than the old result~\rnm~ that at $\Delta=0$
\eqn\nem{S(t;0)=[J_0(2t)]^2}
where $J_0(2t)$ is the Bessel function of order zero.

On the other hand this infinite temperature correlation function
is of great interest because of its relation to the theory of spin
diffusion~\rb-~\rsvw~ which says that in the 
limit $t\rightarrow \infty$  if
spin diffusion is present in the system then  the
asymptotic behavior of $S(t,\Delta)$ should be
\eqn\dif{S(t;\Delta)\sim A t^{-1/2}.}

In the absence of exact results this asymptotic behavior has been
studied by numerical methods for well over 20 years~\rsl-\rfls~.   
By far the most accurate of these numerical studies is the recent work
of Fabricius, L{\"o}w, and Stolze~\rfls~ who have published  
a beautiful high precision  study of the time dependent
two spin correlation functions at various temperatures of the spin 1/2
XXZ antiferromagnet obtained by extrapolating to the thermodynamic limit a
finite size exact diagonalization study of systems of size up to
N=16. From this study the authors were able to give precise estimates
of $S(t;\Delta)$ for $\Delta=1,{\rm cos} (.3 \pi)$ for the time
interval $0 \leq t \leq 4.95$ and they concluded that no
evidence of the asymptotic behavior \dif~ could be seen. This
supports the conjecture made by the present author~\rmc~and by
others~\rczp~ that 
the integrability of the XXZ chain forbids the spin diffusion
asymptotic behavior ~\dif~ and agrees with the absence of diffusion
at low temperature  found by 
Narozhny~\rnar~ for  $-1 \leq \Delta \leq 1.$

However, in ~\rfls~there is no positive statement given as to what the
true asymptotic behavior of their data might be. Here we show
that the numerical results of ~\rfls~can be
beautifully fit with the form 
\eqn\form{f(t; d,A,B,\gamma, \Omega, t_0)
=t^{-d}[A+Be^{-\gamma(t-t_0)}\cos \Omega (t-t_0)]}
which was first suggested in ~\rrmp~ with $d=1/2.$

To test the validity of the form~\form~ we have computed 
\eqn\gfun{g(t;d,A)= t^d S(t;\Delta)-A,}
where $S(t;\Delta)$ is given by the numerical values of ~\rfls~which
were generously sent to us in electronic form, and compared the result
with 
\eqn\hfun{h(t;B,\gamma,\Omega, t_0)=Be^{-\gamma(t-t_0)}\cos \Omega (t-t_0).}
The best results of these comparisons are given graphically in
Figs.1 and 2 for $\Delta=1$ and ${\rm cos}(.3\pi).$  
From these comparisons we conclude that for $2.2 \leq
t \leq 4.95 $ the equality 
\eqn\esf{S(t;\Delta)=f(t;d,A,B,\gamma, \Omega,t_0)}
holds to a very high degree of precision where

1) for $\Delta=1$

\eqn\pta{d=.698,~~ A=.245,~~ B=.0581,~~ \gamma=.70,~~ 
\Omega=4.36,~~ t_0=1.90}

2) for $\Delta={\rm cos}(.3\pi)=.587\cdots$

\eqn\ptb{d=.838,~~ A=.21,~~ B=.114,~~ \gamma=.354,~~ 
\Omega=4.06,~~ t_0=1.96}
These results are to be compared with the exact asymptotics of $\Delta=0$
obtained from ~\nem~as
\eqn\exass{S(t;0)\sim {1\over \pi t}\cos^2(2t-{\pi\over 4})={1\over
2\pi t}\left[1+\cos(4t-{\pi\over 2})\right]}
which is of the form ~\form~ with

3) $\Delta=0$

\eqn\ptc{d=1,~~ A=B={1\over 2\pi}= .159\cdots,~~ \gamma=0,~~ \Omega=4,~~
t_0={5\pi\over 8}= 1.963\cdots}
which is plotted in Fig. 3.

In an attempt to provide some estimate of error in the fitting
parameters we plot in Fig. 4 a fit of the case $\Delta=\cos
(.3 \pi)$ obtained with
\eqn\morefit{d=.784,~~A=.196,~~B=.104~~\gamma=.342,~~\Omega=4.06,
t_0=1.96.}
From comparing the fits of Fig. 2 and 4 we conclude that the estimates
of $\Omega$ and $t_0$ are very stable while the exponent $d$ varied by $.045.$
However because the fit which better fits the points with larger $t$
gives the larger value of $t$ we feel that the conclusion that for
$t\rightarrow \infty$  the exponent $d$ for $\Delta=\cos(.3\pi)$ 
is definitely greater the
value it has for $\Delta=1$ is certainly justified.

\newsec{Discussion}
 
The exponent $d$ in ~\pta-\ptc~ is always substantially greater that 1/2.
Thus, if $f(t;d,A,B,\gamma, \Omega,t_0)$ of ~\form~ does indeed represent
the true asymptotic behavior of $S(t;\Delta)$ then the spin diffusion
form ~\dif~is definitely eliminated. This is in agreement with
~\rfls~and is the basis for the claim in the title of this
paper. However, it must always be kept in mind that no study for a
finite time interval can definitively claim to have seen the true
$t\rightarrow \infty$ behavior of a function. This obvious general
statement is of particular importance here because numerical studies
on the (nonintegrable) classical $(S=\infty)$ 
Heisenberg magnet~\rgma-\rglb~show
that while a form like ~\form~held for times up to the order of $50$  
with an exponent $d$ definitely greater than 1/2 that when times up to
100 were considered the exponent $d$ eventually became the spin
diffusion value $d=1/2.$ The belief embodied in the conjecture of the
absence of spin diffusion for an integrable model is that 
there will never be a scale at which
the spin diffusion form ~\dif~sets in.
We further conjecture  for the (nonintegrable) 
XXZ chains of arbitrary spin $S\geq 1$  that
for suitable small times  the form ~\form~will be an excellent
approximation to the autocorrelation function but that there will exist
some time scale such that for times larger than this scale spin
diffusion will set in. This scale should increase as the spin $S$
decreases and become infinite for the integrable case of $S=1/2.$ 
A similar remark holds for
the addition of further nearest neighbor interactions to the XXZ chain
which also destroy the integrability.

We conclude by remarking that since the data of ~\rfls~for $T=\infty$ 
is so very well fit by ~\form~ a similar simple fitting function
must exist for finite $T.$ In addition  further numerical study should
be able to determine the parameters in ~\form~as functions of
$\Delta$ in the entire range $0 \leq \Delta \leq \infty.$
As a guide to such a study we suggest that the parameter $d(\Delta)$
of ~\form~is a monotonic nonincreasing function of $\Delta$ and that
${\rm lim}_{\Delta \rightarrow \infty}d(\Delta)=1/2.$   
\bigskip
{\bf Acknowledgments}

The author is very indebted to Prof. K. Fabricius for 
providing electronic copies
of the data reported in ref. 9 in graphical form and for helpful discussions. 
This work is supported in part by the NSF under DMR9703543.

\vfill                 
\eject
 \centerline{\epsfxsize=6in\epsfbox{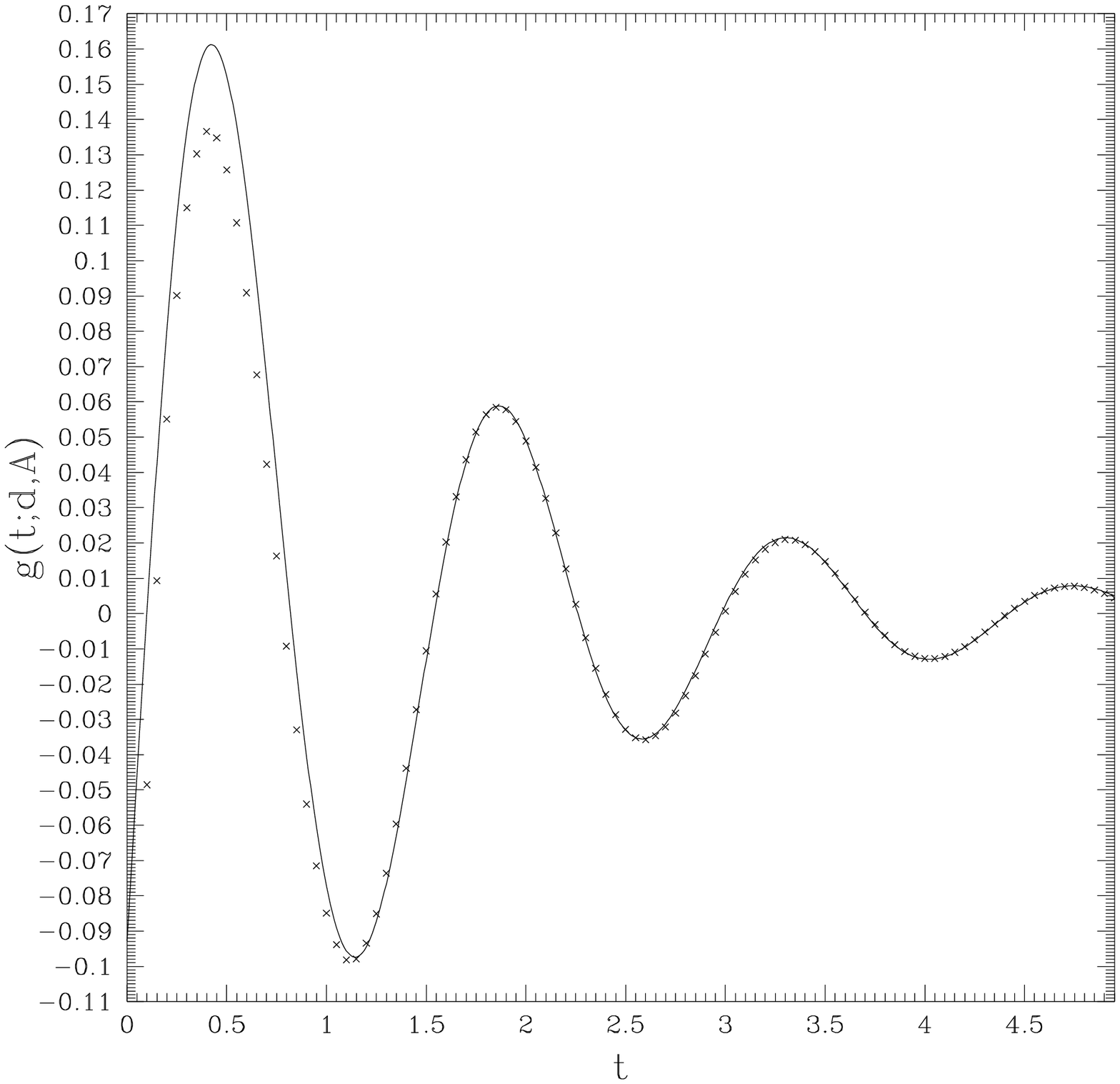}}

Fig. 1 Comparison of the numerical data of ~\rfls~for $\Delta=1$ 
with the form ~\form~with $d=.698,~A=.245,~B=.0581,~\gamma=.7,~
\Omega=4.36\cdots,
~t_0=1.90.$ The points are the function $g(t;,d,A)$
~\gfun~obtained from the data of ~\rfls. The continuous curve is
$h(t;B,\gamma,\Omega, t_0)$~of ~\hfun~with
$B=.0581,~\gamma=.70,~\Omega=4.36,~t_0=1.90.$

\vfill                 
\eject
 \centerline{\epsfxsize=6in\epsfbox{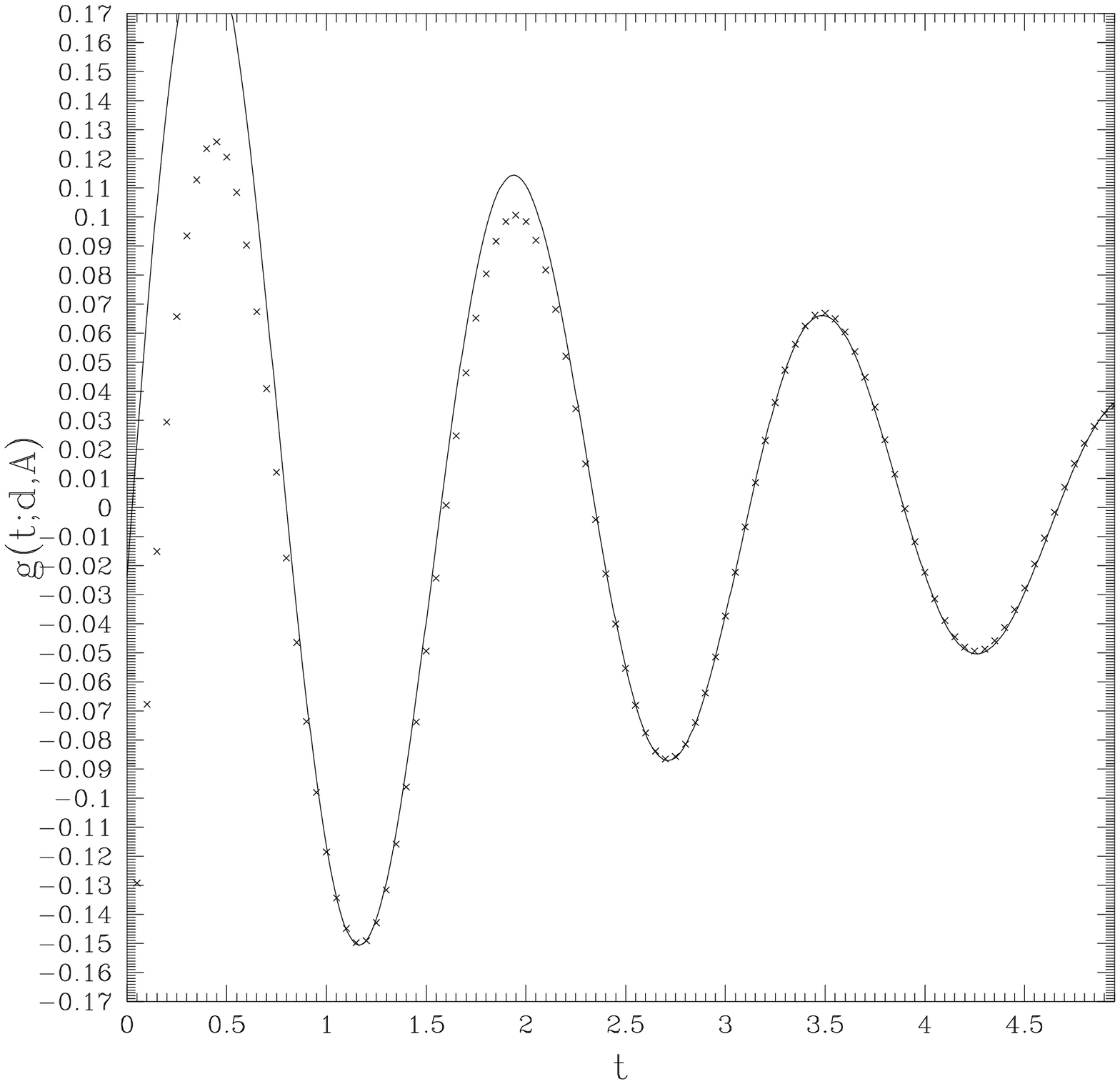}}

Fig. 2  Comparison of the numerical data of ~\rfls~for 
$\Delta=\cos(.3\pi)$ 
with the form ~\form~with $d=.838,~A=.210,~B=.114,~\gamma=.354,~
\Omega=4.06,~t_0=1.96.$ 
The points are the function $g(t;,d,A)$
~\gfun~obtained from the data of ~\rfls. The continuous curve is
$h(t;B,\gamma,\Omega, t_0)$~of ~\hfun~with
$B=.114,~\gamma=.354,~\Omega=4.06,~t_0=1.96.$

\vfill                 
\eject
 \centerline{\epsfxsize=6in\epsfbox{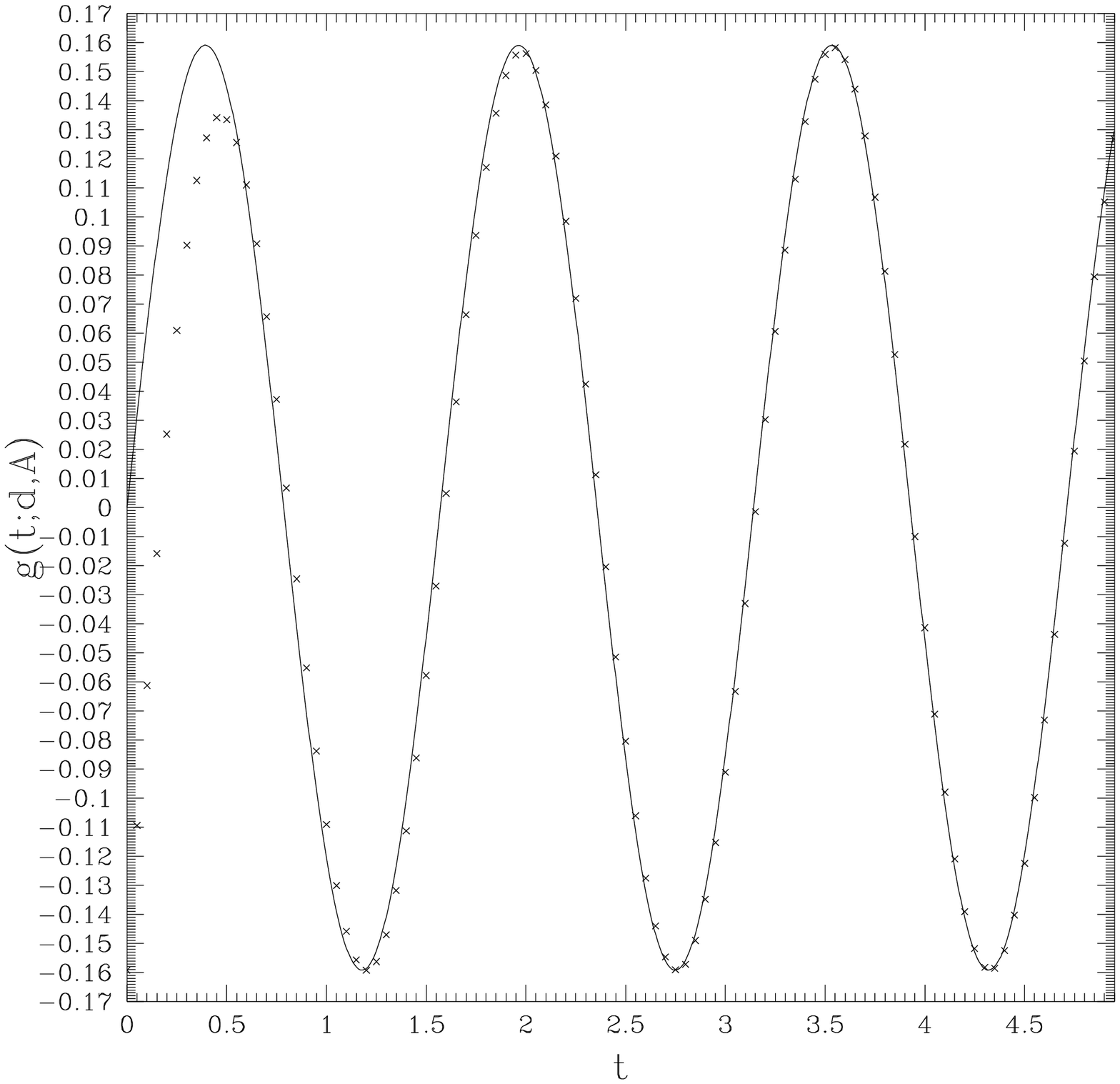}}

Fig. 3  Comparison of the exact result~\nem~for 
$\Delta=0$ 
with the form ~\form~with $d=1.0,~A=B={1\over 2\pi}=.159\cdots,~\gamma=0,~
\Omega=4.0,~t_0={5\pi\over 8}=1.963\cdots.$ 
The points are the function $g(t;,d,A)$
~\gfun~obtained from ~\nem. The continuous curve is
$h(t;B,\gamma,\Omega, t_0)$~of ~\hfun~with
$B={1\over 2\pi}=.159\cdots,~\gamma=0.0,~\Omega=4.0,~t_0={5\pi\over
8}=1.963\cdots.$

\vfill                 
\eject
 \centerline{\epsfxsize=6in\epsfbox{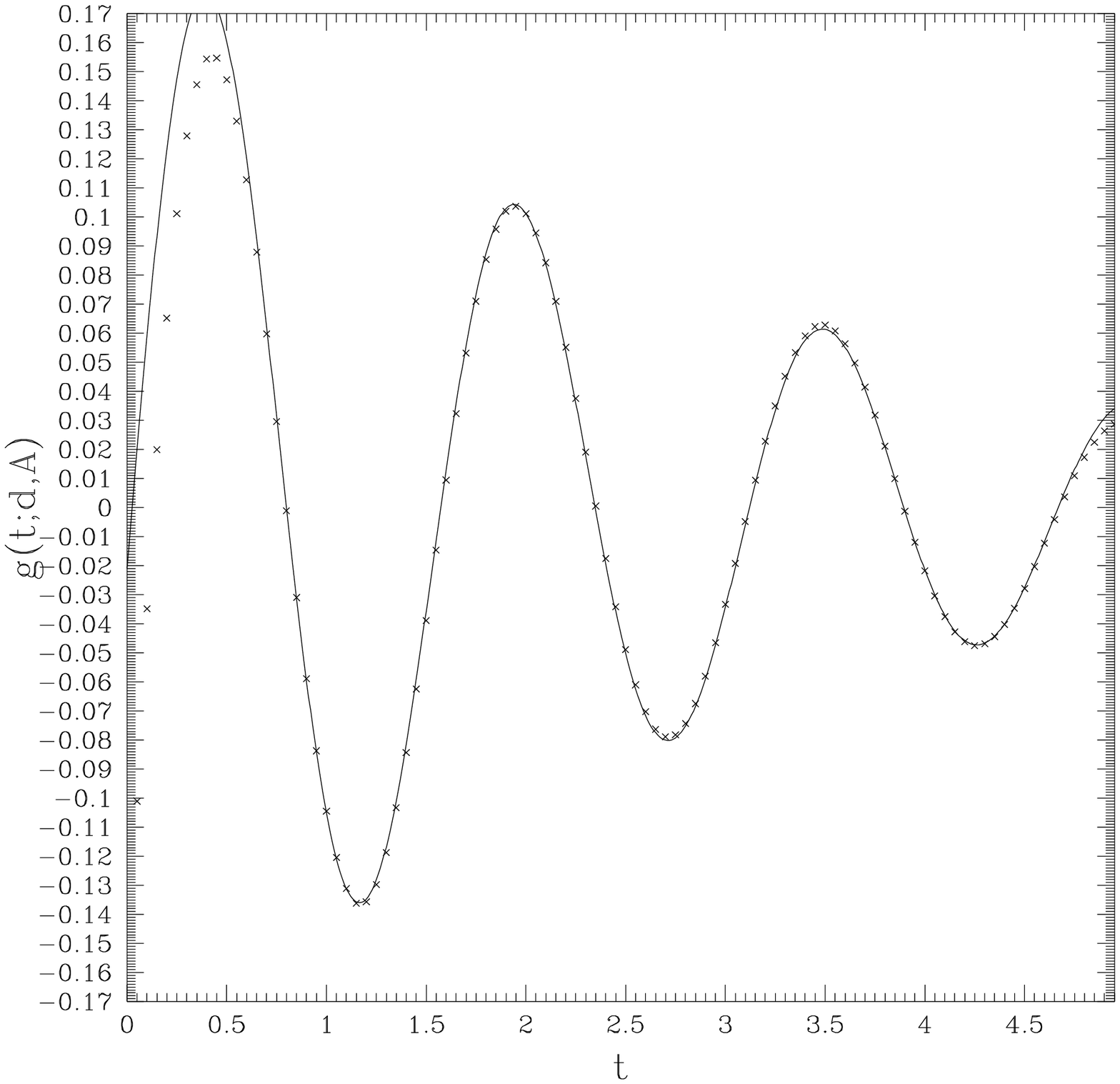}}

Fig. 4  Comparison of the numerical data of ~\rfls~for 
$\Delta=\cos(.3\pi)$ 
with the form ~\form~with $d=.784,~A=.196,~B=.104,~\gamma=.342,~
\Omega=4.06,~t_0=1.96.$ 
The points are the function $g(t;,d,A)$
~\gfun~obtained from the data of ~\rfls. The continuous curve is
$h(t;B,\gamma,\Omega, t_0)$~of ~\hfun~with
$B=.104,~\gamma=.342,~\Omega=4.06,~t_0=1.96.$

\vfill                 
\eject

\listrefs

\vfill\eject

\bye
\end